# Deep-ultraviolet Cherenkov radiation in all-normal-dispersion waveguide enabled by spatial-temporal dynamics

Tiandao Chen[1], Zhiyuan Huang[1,2,3,*], Jinyu Pan[1], Donghan Liu[1,2,3], Ruochen Yin[2,3], Xinglin Zeng[2,3], Jinxin Zhan[2,3], Jiapeng Huang[2,3], Wenbin He[2,3], Xin Jiang[2,3], Hao Hong[4], Kaihui Liu[4], Yuxin Leng[1,5,*], Ruxin Li[1,6] and Meng Pang[1,2,3,5,*]

[1]State Key Laboratory of Ultra-intense Laser Science and Technology, Shanghai Institute of Optics and Fine Mechanics (SIOM), Chinese Academy of Sciences (CAS), Shanghai 201800, China

[2]Russell Centre for Advanced Lightwave Science, Shanghai Institute of Optics and Fine Mechanics (SIOM) and Hangzhou Institute of Optics and Fine Mechanics (HIOM), Hangzhou 311400, China

[3]Zhejiang Key Laboratory of Microstructured Specialty Optical Fiber, Hangzhou Institute of Optics and Fine Mechanics (HIOM), Hangzhou 311400, China

[4]State Key Lab for Mesoscopic Physics and Frontiers Science Center for Nano-optoelectronics, School of Physics, Peking University, Beijing, China

[5]Hangzhou Institute for Advanced Study, University of Chinese Academy of Sciences, Hangzhou 310024, China

[6]Zhangjiang Laboratory, Shanghai 201210, China

*Corresponding author: huangzhiyuan@siom.ac.cn; lengyuxin@siom.ac.cn; pangmeng@siom.ac.cn

**Nonlinear propagation of ultrashort pulses in multi-mode waveguides, featuring complex spatial-temporal dynamics, provides new degrees of freedom in the fields of nonlinear optics and ultrafast lasers. Here, we demonstrate a new scheme of ultraviolet Cherenkov (dispersive-wave) radiation in a gas-filled capillary with unprecedently-high pulse energy, enabled by spatial-temporal dynamics. We found that mJ-level, 40-fs pulses, launched into a large-core capillary filled with high-pressure noble gas, would experience self-phase-modulation and self-steepening effects in this normal-dispersion waveguide, leading to high-intensity shock wave generation and asymmetric spectral broadening. Spatial-temporal dynamics, stemming from strong nonlinear inter-mode coupling, causes spatial shrink and temporal deceleration of the pulse which dramatically alter the capillary dispersion landscape. As a result, a phase-matching point can be created in the ultraviolet, giving rise to the radiation of multi-mode dispersive waves with 100-μJ-level pulse energies and few-fs pulse widths. Our findings inspire new insights into multi-mode nonlinear optics, and the demonstrated high-energy ultraviolet light source with broadband tunability and compact set-up configuration, may find a few applications in time-resolved spectroscopy, ultrafast electronics and femtosecond chemistry.**

Advances in exploring and controlling transient dynamics of matters have been largely contingent upon the development of ultrafast light sources[1-9], with especial focus on their photon-energy and temporal-width performance. Ultrafast light sources in the deep and vacuum ultraviolet (100-300 nm), with unique advantages of high photon energy and short optical wavelength, have garnered extensive attentions in the fields of nonlinear optics and ultrafast science over the past decades[10-29]. While the crystal lifetime and laser damage due to strong material absorption, remain major challenges of crystal-based ultrafast light sources at deep and vacuum ultraviolet wavelengths[10-12], the scheme of harmonic generation using gas jet at these wavelengths normally has relatively-low conversion efficiency and limited tunability[15, 16].

Dispersive-wave emission, also known as Cherenkov radiation[30,31], has been widely regarded as an important strategy of high-efficiency conversion of ultrafast laser pulses to short wavelengths. In previous literatures[17-29], dispersive-wave emission has always been considered as an accompanying phenomenon that is closely associated with soliton dynamics. In a generic picture, the high-order soliton, launched into an optical waveguide with anomalous dispersion, would first experience soliton self-compression and dramatic spectral broadening. In the presence of high-order dispersion, a phase-matched energy transfer from the pump (in anomalous-dispersion side) to the dispersive-wave wavelength (in the normal-dispersion side) gives rise to self-seeded optical Cherenkov radiation via the cascaded four-wave-mixing process[30,31]. While this canonical physics manifests itself in a variety of nonlinear-optics platforms including solid-core or hollow-core micro-structured fiber[17,18], gas-filled capillary waveguides[19-29] and photonic chip-scale devices[32,33], the power-scaling law of soliton dynamics[19,34] strictly defines the relationship between the radiated pulse energy and the system geometric dimension (diameter and length of the waveguide).

In recent several years, spatial-temporal dynamics in multi-mode waveguides, providing additional degrees of freedom for investigation and control of light propagation, have attracted significant interest in the society of optics[35-45]. The introduction of this concept has also opened up a rich world of multi-mode nonlinear optical systems with remarkable diversity and complexity[42], inspiring unprecedented possibilities in nonlinear-optics experiments[43-45]. In this *Letter*, we unveiled the formation of a new type of phase-matching for optical Cherenkov radiation, stemmed from strong nonlinear inter-mode coupling. We demonstrate that the optical spectrum of a mJ-level ultrafast pulse, when propagating in a normal-dispersion capillary filled with high-pressure noble gas, would be asymmetrically broadened due to self-phase-modulation (SPM) and self-steepening (SST) effects, generating in the time domain a packet of shock wave with high peak intensity and few-femtosecond pulse width. The enhanced nonlinearity would result in strong nonlinear coupling between capillary optical modes, effectively decelerating the shock-wave pulse in longitudinal direction. This largely-reduced pulse group velocity creates new phase matching with the quasi-linear light component in the deep ultraviolet. Multi-mode ultraviolet Cherenkov (dispersive-wave) radiation can therefore be obtained at the capillary output, with up to 100-μJ pulse energy and excellent spectral coherence, supporting a compressed pulse duration of <2 fs.

The multi-mode Cherenkov radiation (MMCR) process is conceptually illustrated in Fig. 1a. When a large-core (500-μm-diameter) capillary fiber is filled with high-pressure (9.3 bar in the experiment) Ne gas, the gas-material dispersion mainly determines the dispersion property of the capillary waveguide, whose second-order term remains to be normal ($\beta_2 > 0$) over the wavelength range from 200 nm to 1000 nm. The pump pulse, launched into the fundamental optical mode of the capillary waveguide, is at ~800 nm with a ~40-fs pulse width and a mJ-level pulse energy. During the propagation, the ultrafast pump pulse first experiences strong SPM and SST effects[40,41,46,47] due to relatively-high gas pressure and therefore strong waveguide nonlinearity, leading to the asymmetrical spectral broadening[40,46,47] in the frequency domain and the generation of a packet of ultrashort shock wave[40,46,47] in the time domain. The appearance of this few-femtosecond shock wave further enhances the optical peak-intensity and nonlinearity, giving rise to strong inter-mode energy coupling from the fundamental to several high-order modes.

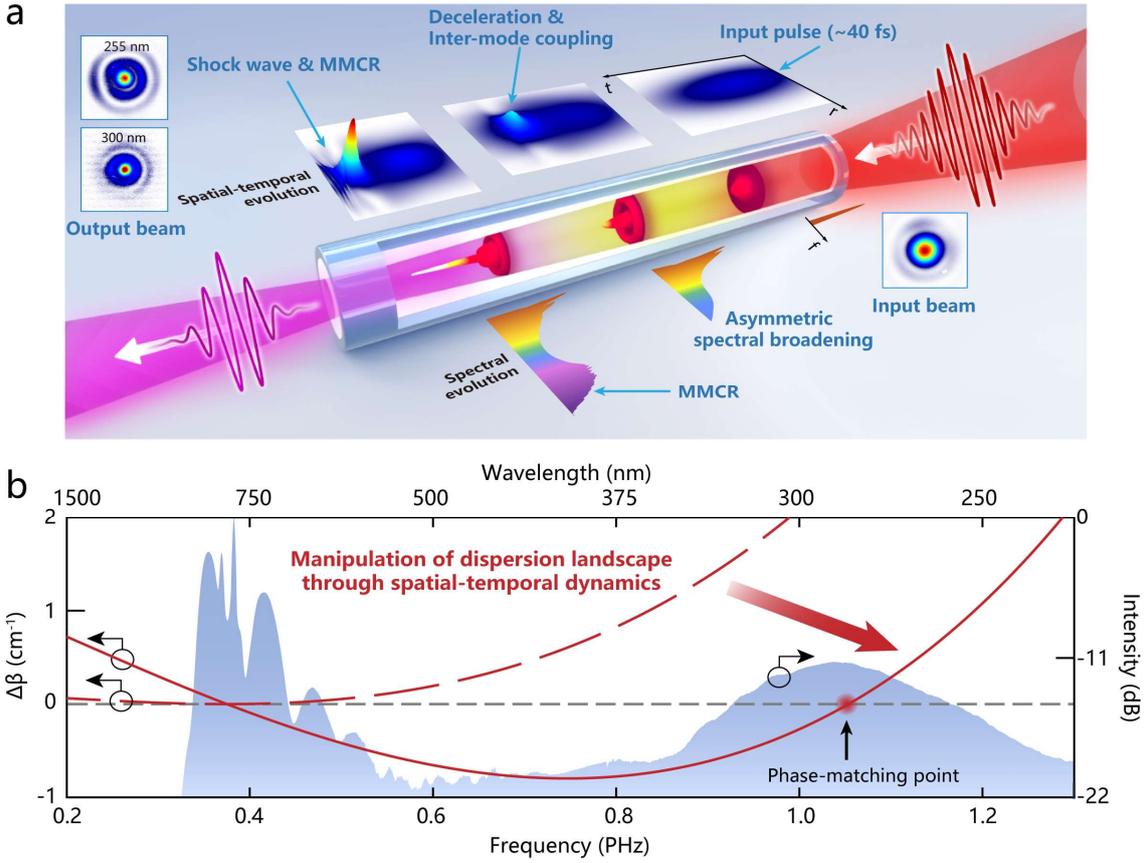

**Fig. 1 | Conceptual illustration of the MMCR generation mechanism and experimental results. a**, Schematic of MMCR generation mechanism in a Ne-filled HCF. The top panel illustrates the dynamics of inter-mode coupling, pulse deceleration, shock wave formation, and MMCR, with the corresponding spectral evolution depicted below. The beam profile of the focused input pump and near-field images of the UV MMCR pulses at different output wavelength, measured using a CCD camera, as shown at the input and output ends, respectively. **b**, Dephasing rate $\Delta\beta$ from Eq. (1) between the 800-nm pump and MMCR for a 9.3-bar Ne-filled HCF, calculated without (red dashed line) and with (red line) the $\beta_{STD}$ of 0.6 fs/cm. Phase matching is achieved at the zero-dephasing point (red dot). The blue-shaded area represents the measured spectrum under pumping conditions of 565-μJ pulse energy and ~40-fs pulse duration. The HCF used in the experiment has a length of 40 cm and a core diameter of 500 μm, and is filled with 9.3-bar Ne gas.

As illustrated in Fig. 1a, such a spatial-temporal evolution of the pump pulse results in a spatial-profile shrink (similar as the self-focusing effect), which causes the reduction of the pulse group velocity along the capillary longitudinal direction. This pulse deceleration due to spatial-temporal dynamics varies the dispersion landscape of the waveguide (see Fig. 1b) through introducing an additional group-velocity-related term ($\beta_{STD}$), and the phase mismatch between the pump pulse and the optical Cherenkov radiation (some quasi-linear waves) as the function of light frequency ($\omega$) can be expressed as:

$$\Delta\beta = \frac{\beta_2}{2}(\omega-\omega_0)^2 + o[(\omega-\omega_0)^2] - \beta_{STD}(\omega-\omega_0) \qquad (1)$$

Where $\Delta\beta$ represents the dephasing rate between the two waves, $\beta_2$ the second-order dispersion at the pump wavelength $\omega_0$, $o$ all the higher-order ($\geq 3$) dispersion terms. $\beta_{STD}$ in Eq. (1) represents the propagation constant variation of the shock wave due to the nonlinear spatial-temporal dynamics, which can be expressed as $\beta_{STD} = \beta_1(\theta^2/2)$, where $\beta_1$ is the first-order term of the propagation constant and $\theta$ the effective divergence angle due to the pulse spatial shrink, see more details in Supplementary note 1.

As illustrated in Fig. 1b, the introduction of $\beta_{STD}$ term can strongly modulate the otherwise parabolic-shape dispersion curve (see red dashed line Fig. 1b) of the waveguide, creating a phase-matching ($\Delta\beta = 0$) point in the high-frequency (shorter-wavelength) side. When the energy of 40-fs pump pulse launched into the capillary is 565 µJ, $\beta_{STD}$ can be estimated to be ~0.6 fs/cm (see Supplementary notes 1 and 2), giving rise to a phase-matched wavelength at ~290 nm, see Fig. 1b. In the experiment, strong Cherenkov radiation at ~290 nm was observed from the 40-cm-long capillary (see the blue shallow area in Fig. 1b), exhibiting good agreement with the theoretical prediction. Note that in this large-core capillary filled with high-pressure Ne gas, the contribution from gas material dominates the waveguide dispersion, with trivial mode-related contribution. Thus, light components at several low-order optical modes of the capillary waveguide experiences almost the same dispersion landscape (see red lines in Fig. 1b).

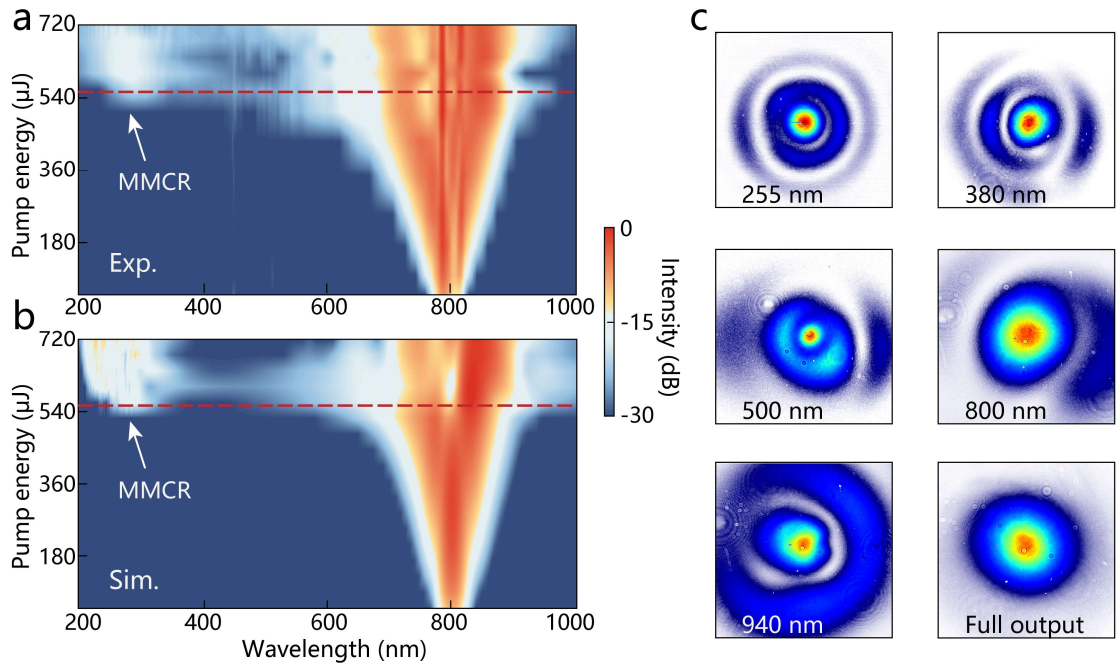

**Fig. 2 | Experimental and numerical results of spectral evolution with pump energy and spatial profiles of the output spectrum. a**, Measured output spectra from the HCF as a function of pump pulse energy. The system parameters are the same as Fig. 1b. **b**, Corresponding numerical simulations. The red dashed line in (**a**) marks the spectrum shown in Fig. 1b, while the red dashed line in (**b**) indicates the case corresponding to Fig. 3. **c**, Measured near-field beam profiles at different wavelengths for the spectrum shown in Fig. 1b.

As described in Eq. (1), the phase-matching wavelength of the MMCR can be tuned through varying the value of $\beta_{STD}$ which, in practice, can be achieved simply through adjusting the pump pulse energy. This wavelength tunability was experimentally realized, see Fig. 2a, Methods and Supplementary note 4 for more experimental details. As illustrated in Fig. 2a, when we gradually increased the energy of the ~40-fs pump pulse from 50 µJ to 720 µJ, we recorded the output spectra from the capillary at different pump energies. It can be found that at pump energies below 450 μJ, the symmetrical spectral broadening could be mainly attributed to the SPM effect, while at pump energies above 450 μJ obvious asymmetry of the broadened spectrum due to the self-steepening effect[47] could be observed, leading to the fact that more light components were generated at shorter wavelengths. These shorter-wavelength components could work as the seed light of the Cherenkov radiation, only if their phase-matching with the high-intensity driving pulse was satisfied.

Such phase-matching and therefore high-efficiency Cherenkov radiation were obtained in the experiment at pump pulse energies of >520 μJ. As illustrated in Fig. 2a, at 520 μJ pump energy the radiation started to appear at ~300 nm, and its wavelength shifted gradually to ~250 nm as the pump energy gradually increased to 720 μJ. These experimental observations can be perfectly reproduced in our numerical simulations using multi-mode unidirectional pulse propagation equation (MM-UPPE)[19,48,49], see Fig. 2b and more details in Supplementary note 3. At a pump pulse energy of 565 μJ, the output optical spectrum (marked by the red dashed line in Fig. 2a) was carefully measured in experiments, and the measured near-field beam profiles at five different wavelengths, together with that of the full output, are illustrated in Fig. 2c, verifying further that this MMCR process involves several optical modes of the capillary and therefore is inherently spatial-temporal dynamics. As this pump level, the pulse energy of the deep-ultraviolet radiation was measured to be ~18 μJ, corresponding a conversion efficiency of ~3.2%.

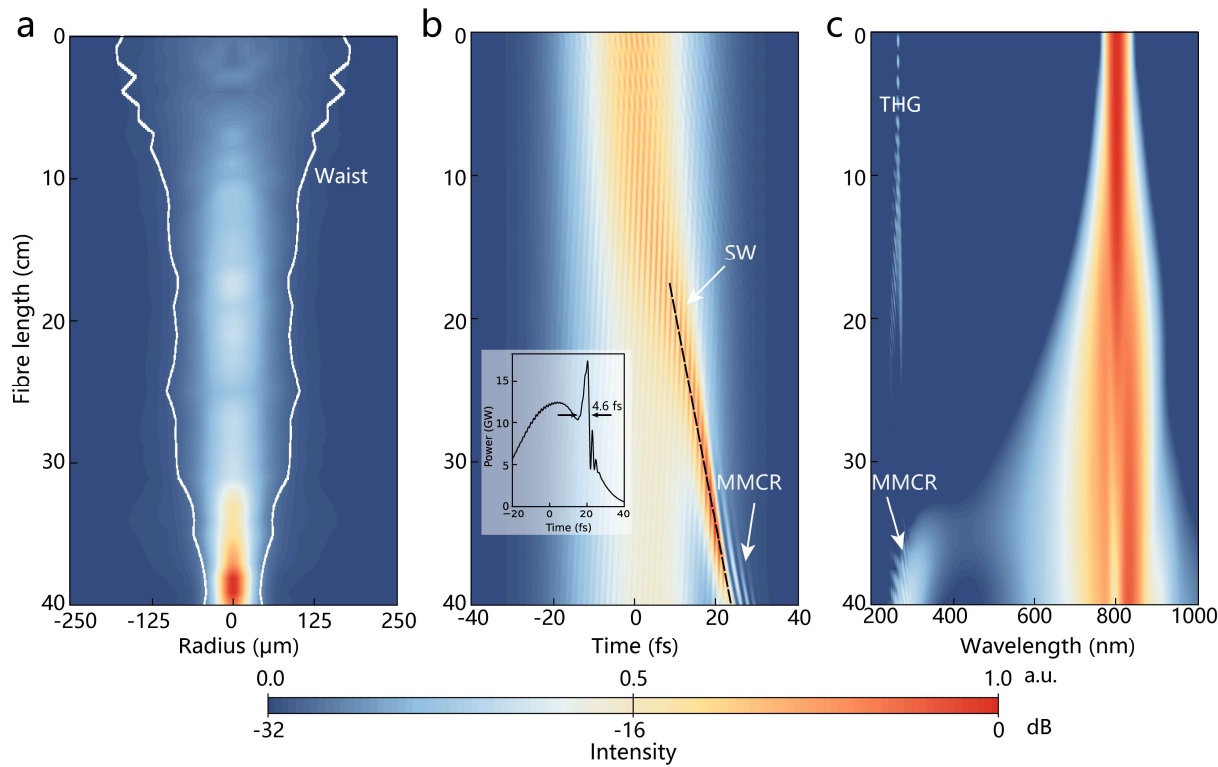

**Fig. 3 | Simulated evolution dynamics of the MMCR process along the fibre length for the case marked in Fig. 2b.** Simulated spatial (**a**), temporal (**b**) and spectral (**c**) evolution of the 565-μJ pump pulse as a function of fibre length. The white line in **a** indicates the beam waist at the temporal peak. The inset in **b** shows the pulse profile at 35 cm. SW, shock wave.

To better illustrate the spatial-temporal dynamics of pulse propagation in the capillary, we present the 3-demensional nonlinear evolution of the 565 μJ pulse over the 40-cm capillary (see Fig. 3), using the simulation data. As illustrated in Fig. 3a, the optical beam waist gradually decreases as the pump pulse propagates in the capillary, and the shrink of beam spatial profile, corresponding the nonlinear inter-mode coupling, is enhanced by the generation of high-intensity shock wave in the time domain (see Fig. 3b), on one hand. This spatial beam shrink, on the other hand, results in the deceleration of the shock-wave pulse which can be clearly observed in Fig. 3b. In the spectral domain (see Fig. 3c), the generation of shock wave due to self-steepening effects causes directly the asymmetric spectral broadening (biased towards short wavelengths), and consequently results in strong optical Cherenkov radiation to the phase-matched wavelength (~290 nm).

The 3-dimensional pulse evolution could be inspiringly illustrated using spatial-temporal and spectral-temporal diagrams, see Fig. 4 and Supplementary Movie 1. As illustrated in Figs. 4a–4c, the generation of shock wave at ~30 cm enhances the spatial shrink (nonlinear mode-coupling) effect of the optical beam, further increasing the optical peak intensity from 148 TW/cm$^2$ to 362 TW/cm$^2$. As illustrated in the spectral-temporal diagram (see Figs. 4d–4f), the asymmetric spectral broadening results mainly from the generation of shock wave, providing the seed light of the MMCR. The strong MMCR at the ultraviolet wavelength comes from the temporally-decelerated optical shock wave, see Fig. 4f.

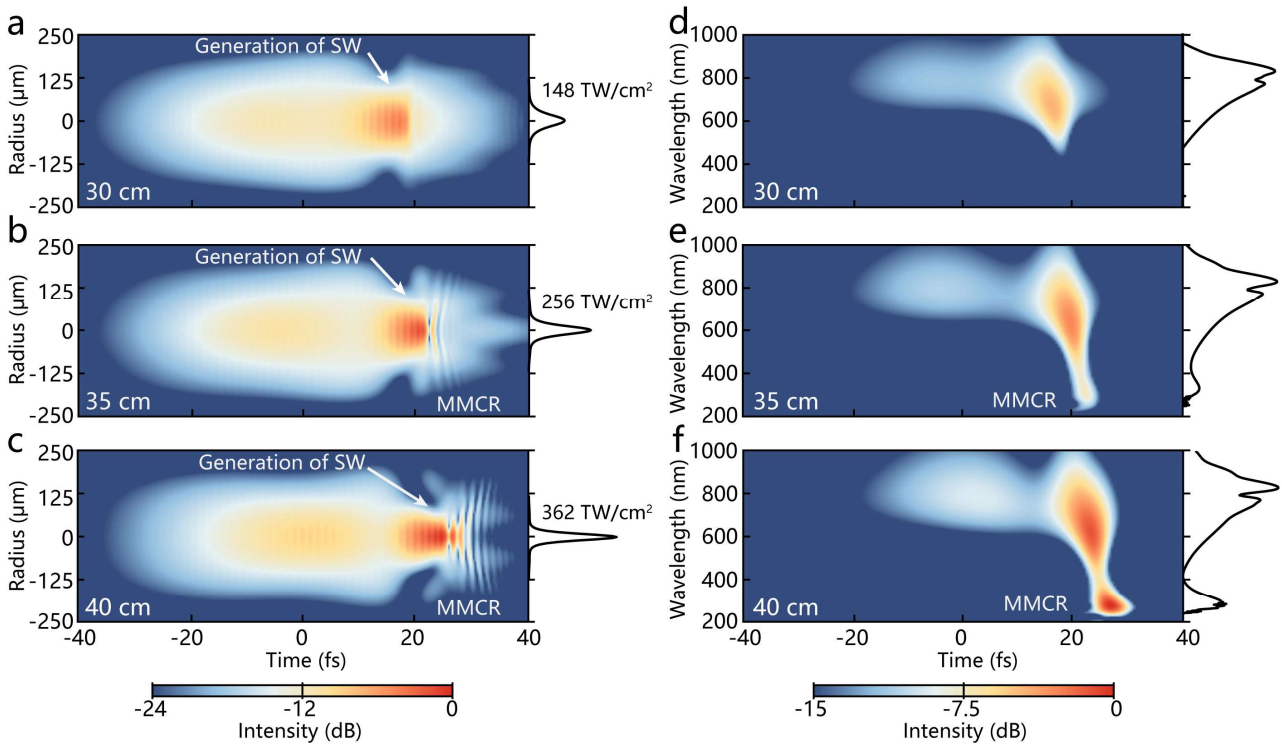

**Fig. 4 | The spatial-temporal distribution and on-axis spectral-temporal diagrams (spectrograms) at different fibre positions for the case in Fig. 3. a-c**, Spatial-temporal distribution. The black lines indicate the radial intensity distributions at the pulse peak. **d-f**, On-axis spectrogram analysis. The black lines indicate the pulse spectra.

We further investigated the mode composition, coherence and phase characteristics of the MMCR pulse[17,50], and found that the radiated UV pulse is mainly composed of four low-order capillary optical modes (containing >95% of the total energy). Due to the perfect radial symmetry of the system, all the four ($HE_{11}$, $HE_{12}$, $HE_{13}$ and $HE_{14}$) capillary modes under study also exhibit radial symmetry, see Fig. 5. The simulation results (see Fig. 5a) illustrate that even though the radiated UV pulses at different optical modes have different spectral shapes, all of them exhibit good coherence over the whole spectral range. In the time domain, as illustrated in Fig. 5b the MMCR pulses at different modes have trivial inter-mode walk-off owing to the weak mode-dependent dispersion in the large-core capillary. The pulse chirp at the capillary output was also found to be smooth, and could be acceptably eliminated through compensating only the second-order term. As illustrated in Fig. 5c, through introducing a group-delay-dispersion value of $-1.8$ fs$^2$, the MMCR pulse could be compressed from 4.4 fs to 1.9 fs, which is already quite close to its Fourier-transform-limit value of 1.5 fs.

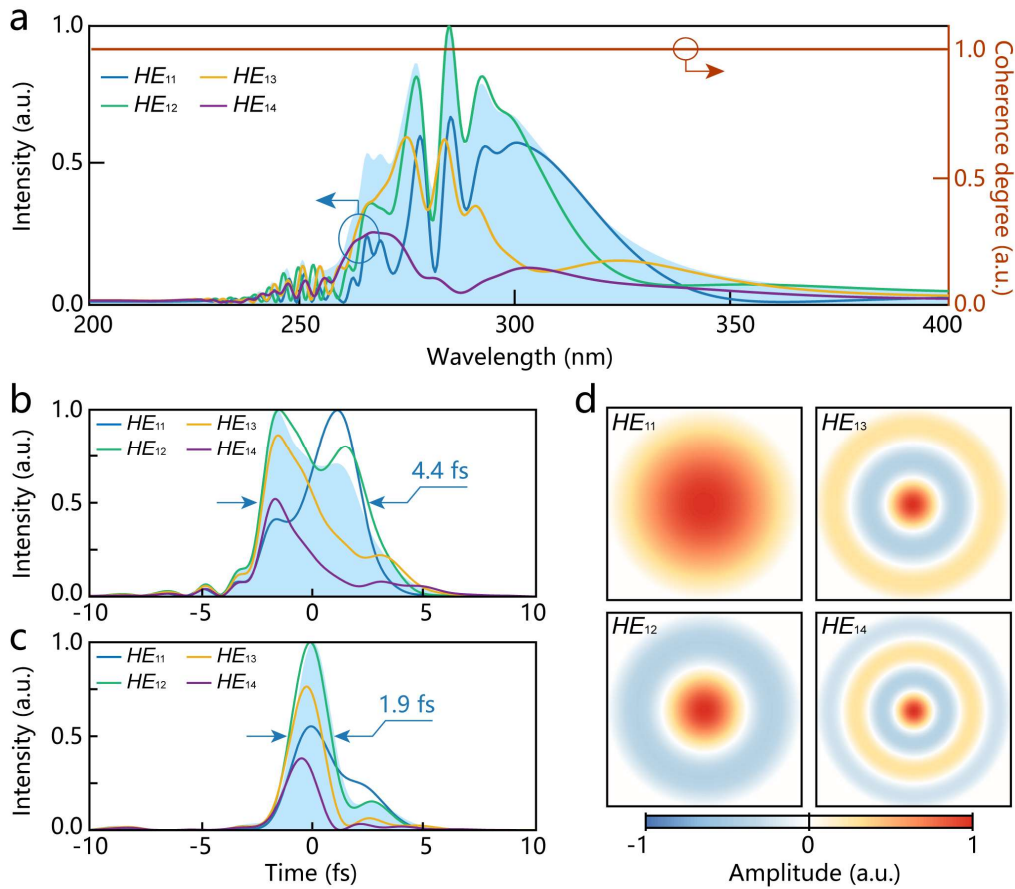

**Fig. 5 | Mode composition, coherence and phase characteristics of the multimode UV pulse generated via MMCR. a**, Simulated spectra for individual modes (colored lines) and the complex degree of first-order coherence (orange line) of the multimode UV pulse at the HCF output in Fig. 3. **b**, Corresponding temporal envelopes for each mode. **c**, Compressed pulse obtained by applying -1.8 fs$^2$ group delay dispersion to the output multimode UV pulse. The blue shadows in (**a-c**) represent the superposition of UV pulses across all modes. **d**, Normalized transverse amplitude distributions of the $HE_{11}$ to $HE_{14}$ modes.

In experiments, we found that the MMCR phenomenon, induced by the nonlinear spatial-temporal dynamics and the subsequent dispersion manipulation, is generic, and can be observed over a wide range of system parameters, see Fig. 6. While using the same capillary, the phase-matched MMCR wavelength can be tuned from ~300 nm to ~250 nm through simply varying the pump pulse energy (see Figs. 6a and 6b), the pulse energy of the MMCR could be scaled up and down through using different system parameters, including the pump pulse energy, the capillary geometric dimension as well as the gas type and pressure filled in the capillary. Two typical examples are illustrated in Figs. 6c and 6d. When the capillary with the same geometric dimension (500-μm core diameter and 40-cm capillary length) was filled with 3 bar Ar gas, the 40-fs pump pulse with a pulse energy of 175 μJ could also give rise to 7.7-µJ MMCR pulse at ~300 nm, corresponding to a conversion efficiency of ~4.4%, see Fig. 6c. The energy scaling-up is also possible when a capillary with a larger core diameter is used. As illustrated in Fig. 6d, the use of 800 µm-core-diameter, 1-m-length capillary filled with 6.5-bar He gas, can support the generation of MMCR pulse generation with a pulse energy of 113 µJ at ~300 nm (corresponding to a conversion efficiency of ~5.1%), when pumped by the 40-fs pulse with an energy of 2.2 mJ. See also more results in Supplementary note 5.

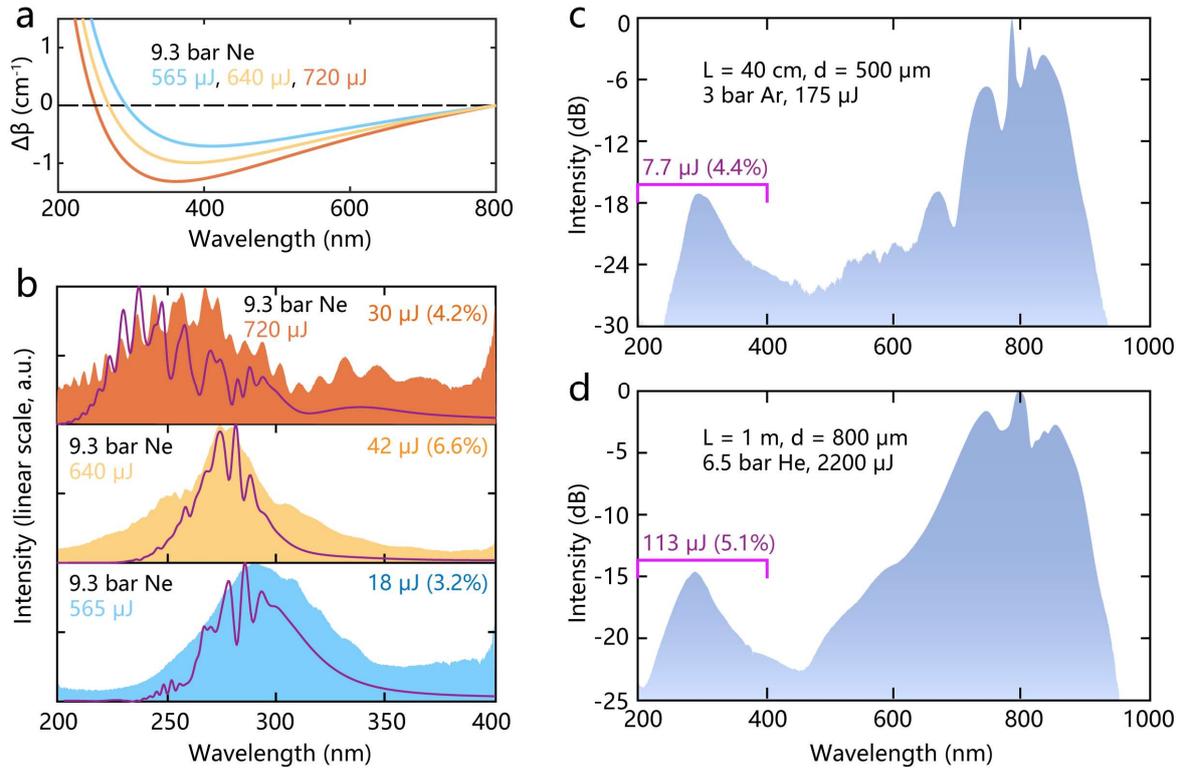

**Fig. 6 | Central-wavelength tuning and energy scaling of MMCR. a**, Phase-matching diagram of MMCR at different input energies. **b**, Corresponding UV pulse spectra at different phase-matching wavelengths. The pump pulse duration is ~40 fs, and the HCF has a length of 40 cm and a core diameter of 500 µm. Simulated multimode UV spectra (red lines) are obtained by adjusting the input energy within a 10% range to match the measurements. **c**, Measured spectrum for a 175-µJ, ~40-fs pump in a 40-cm-long, 500-µm-core filled with 3-bar Ar gas. **d**, Measured spectrum for a 2.2-mJ, ~40-fs pump in a 1-m-long, 800-µm-core filled with 6.5-bar He gas.

Compared with conventional optical Cherenkov radiation, based on soliton dynamics, in gas-filled capillary, the MMCR mechanism demonstrated here can efficiently scale up the output ultraviolet pulse energy, and more importantly it can largely simplify the system set-up which is critical for applications. First, the pump pulse width is at 40-fs level (rather than 10-fs level) which can be easily obtained from widely-used Ti: sapphire ultrafast laser systems. Second, in the conventional dispersive-wave-radiation scheme, the scaling-law of soliton dynamics always asks for a lower gas pressure at a larger capillary core diameter, so as to ensure a balance between the dispersion contributions from the filled-gas material and the waveguide geometry to obtain anomalous dispersion at the pump wavelength and simultaneously to select the right phase-matched Cherenkov-radiation wavelength[19,34]. The low gas-pressure at large waveguide size would result in weak optical nonlinearity and therefore long nonlinear length (corresponding to large system geometry), which remains a major challenge of capillary systems for practical applications. Simple calculations exhibit that in the conventional scheme[19], a capillary length of ~20 m will be necessary for obtaining 100-µJ-level dispersive-wave pulse at ~300 nm, when the core diameter of the capillary is 800 µm and the 10-fs pulse width is required for the mJ-level pump pulse, see more details in Supplementary note 6. In contrast, in this MMCR scheme demonstrated here 40-fs, mJ-level pump pulse can efficiently drive the ~300-nm MMCR with a UV pulse energy of 113 µJ, using merely one stage of 1-m-length, 800-µm-diameter capillary, see Fig. 6d.

The phenomena of optical Cherenkov radiation in multi-mode waveguides have attracted considerable attentions in recent several years[35,36,40,41]. While multi-mode dispersive-wave radiation, originating from soliton dynamics in graded-index multi-mode fibre[35,36], has been observed with the comb-like, broadband spectrum, the spatial-temporal dynamics and inter-mode phase-matching have also been explored in solid-core multi-mode fibre or bulk-material nonlinear medium for generating few-µJ-level visible supercontinuum light[40,41,51]. The extension of such spatial-temporal mechanisms to the platform of gas-filled, all-normal-dispersion capillary waveguide, enabling the generation of high-energy (100-µJ-level) ultraviolet dispersive-wave pulses with high optical coherence, has not be explored to date.

In summary, we have shown that the nonlinear spatial-temporal pulse propagation in a large-core capillary waveguide, could decouple the phase-matched optical Cherenkov radiation from the soliton pulse compression, largely improving both the energy-scalding capability and the set-up compactness of the system. From a nonlinear physics perspective, our results reveal a further example of the great utility of multi-mode nonlinear optics systems, shedding some new insights into complex spatial-temporal pulse dynamics. Form the application viewpoint, the demonstrated ultraviolet-tunable light source with remarkably-high pulse energy, few-femtosecond pulse width and vastly-simple laser set-up may render great application potential in advanced photolithography, ultrafast electron microscopy and femtosecond optics experiments.

## Methods

**Experimental set-up.** The system was driven by a commercial Ti: sapphire femtosecond laser delivering 800 nm, ~40 fs pulses at a repetition rate of 1 kHz. Pump pulses with energies up to 3.5 mJ were launched into HCFs with core diameters of 500 µm or 800 µm. See Supplementary note 4 for more details of the set-up.

**Pulse spectrum and energy measurements.** Pulse spectrum were measured using two spectrometers: one for the UV range (Maya2000-Pro, 200-400 nm, Ocean Insights) and one for the visible (VIS) to near-infrared (NIR) range (Maya2000-Pro, 210-1100 nm, Ocean Optics). Both were calibrated. The full optical spectrum at the HCF output (Figs. 1b, 2a, 6b-d) was obtained by combining data from both spectrometers and interpolating across the overlapping region. For these measurements, the output beam was split by a fused silica wedge. The transmitted beam was directed into an integrating sphere coupled to the UV spectrometer, and the reflected beam into another integrating sphere coupled to the VIS-NIR spectrometer.

Input pulse energy was measured with a thermal power meter (3A-P, Ophir). The energy of the multimode UV pulses was measured using a combination of bandpass filters (Pelham Research Optical for 255 nm; FGUV11, Thorlabs for 275-375 nm), a photodiode (PD300-UV, Ophir), and the UV spectrometer. As an example, for the broadband UV pulse generated at 565-µJ pump energy and 9.3-bar Ne, a bandpass filter first isolated a spectral portion. Its energy and spectral intensity were measured with the photodiode and UV spectrometer to calibrate the spectral energy density. The UV spectrometer then measured the spectral intensity of the unfiltered multimode UV pulse. Using the calibrated spectral energy density, the total energy of the broadband UV pulse was estimated to be 18 µJ. The same method was used for UV pulses with different bandwidths and central wavelengths (results in Figs. 5b-d).

**Beam profile measurements.** Near-field beam profiles of the output spectra (Figs. 1a and 2c) were measured at different wavelengths using optical filters (Pelham Research Optical for 255 nm; FBH series, Thorlabs for 300-940 nm) and a CCD camera (BGS-USB3-SP932U, Ophir-Spiricon).

## Data availability

The data generated in this study is available from the corresponding author upon reasonable request.

## Code availability

The code used in this paper is available from the corresponding author upon reasonable request.

## Acknowledgements

This work was supported in part by the National Natural Science Foundation of China (No. W2541021 to M.P., No. 62505330 to J.Y.P., No. 12388102 to R.X.L., No. 42550189 to K.H.L.), the Strategic Priority Research Program of the Chinese Academy of Science (No. XDB0650000 to Z.Y.H. and M.P.), the Shanghai Science and Technology Plan Project Funding (No. 23JC1410100 to Z.Y.H. and M.P.), the National Postdoctoral Program for Innovative Talents (No. BX20250361 to J.Y.P.), the China Postdoctoral Science Foundation (No. 2025M780803 to J.Y.P.), Shanghai Municipal Science and Technology Major Project (Z.Y.H. and M.P.), Fuyang High-level Talent Group Project (J.P.H, W.B.H., X.J. and M.P.).

## Author contributions

Z.Y.H. and M.P. conceived the work. T.D.C and Z.Y.H carried out the experiments. W.B.H., J.P.H., X.J., H.H, K.H.L, Y.X.L. and R.X.L. provided necessary experimental equipment and resources. T.D.C. performed the numerical simulations. T.D.C., Z.Y.H. and M.P. made the theoretical and experimental analysis and wrote the manuscript, Z.Y.H., Y.X.L. and M.P. supervised the project. All authors contributed to the discussion of the results and the editing of the manuscript.

## Competing interests

The authors declare no competing interests.

## Additional information

Supplementary information is available for this paper at xxxxxx.

# Supplementary Information for

# "Deep-ultraviolet Cherenkov radiation in all-normal-dispersion waveguide enabled by spatial-temporal dynamics"

Tiandao Chen[1], Zhiyuan Huang[1,2,3,*], Jinyu Pan[1], Donghan Liu[1,2,3], Ruochen Yin[2,3], Xinglin Zeng[2,3], Jinxin Zhan[2,3], Jiapeng Huang[2,3], Wenbin He[2,3], Xin Jiang[2,3], Hao Hong[4], Kaihui Liu[4], Yuxin Leng[1,5,*], Ruxin Li[1,6] and Meng Pang[1,2,3,5,*]

[1]State Key Laboratory of Ultra-intense Laser Science and Technology, Shanghai Institute of Optics and Fine Mechanics (SIOM), Chinese Academy of Sciences (CAS), Shanghai 201800, China

[2]Russell Centre for Advanced Lightwave Science, Shanghai Institute of Optics and Fine Mechanics (SIOM) and Hangzhou Institute of Optics and Fine Mechanics (HIOM), Hangzhou 311400, China

[3]Zhejiang Key Laboratory of Microstructured Specialty Optical Fiber, Hangzhou Institute of Optics and Fine Mechanics (HIOM), Hangzhou 311400, China

[4]State Key Lab for Mesoscopic Physics and Frontiers Science Center for Nano-optoelectronics, School of Physics, Peking University, Beijing, China

[5]Hangzhou Institute for Advanced Study, University of Chinese Academy of Sciences, Hangzhou 310024, China

[6]Zhangjiang Laboratory, Shanghai 201210, China

*Corresponding author: huangzhiyuan@siom.ac.cn; lengyuxin@siom.ac.cn; pangmeng@siom.ac.cn

# Supplementary note 1: The group-velocity-related dispersion modulation term

As mentioned in the main text, the spatial-profile shrink of the pules inside the HCF is similar as the self-focusing effect. Therefore, we can employ a similar approach to express the group velocity of such pulses. According to Ref. [1], a self-focusing ultrashort pulse propagates with a group velocity $v_g^{SF}$ given by:

$$v_g^{SF} = v_g \cos\theta \tag{S1}$$

where $v_g$ is the group velocity of a plane wave, in this case very close to the speed of light in vacuum $c$. The parameter $\theta$ denotes the ratio of the radial wave vector to the total wave vector, and can be estimated from the divergence angle formula $\theta = \lambda_0/(\pi w_0)$. Here, $\lambda_0$ is the central wavelength of the pulse and $w_0$ is the waist radius at the temporal peak. Since $\beta_1^{SF} = \left(v_g^{SF}\right)^{-1}$, $\beta_1 = \left(v_g\right)^{-1}$ and $w_0 \gg \lambda_0$, Eq. (S1) can be expanded as:

$$\beta_1^{SF} \approx \beta_1 \left(1 + \frac{\theta^2}{2}\right) \tag{S2}$$

Consequently, the dispersion modulation term caused by group velocity deceleration $\beta_{STD} = \beta_1^{SF} - \beta_1$ can be expressed as:

$$\beta_{STD} = \frac{\beta_1}{2}\theta^2 \tag{S3}$$

**Supplementary note 2: Estimation of the dispersion modulation term**

We use the divergence angle formula $\theta = \lambda_0/(\pi w_0)$ to estimate the value of $\beta_{\text{STD}}$. The Eq. (S3) can be rewritten as:

$$\beta_{\text{STD}} = \frac{\beta_1}{2}\left(\frac{\lambda_0}{\pi w_0}\right)^2 \quad (S4)$$

Equation (S4) provides an excellent estimate of the $\beta_{\text{STD}}$. Using the simulation results shown in Fig. 3a of the main text, the waist radius at the temporal peak near the multi-mode Cherenkov radiation (MMCR) point is ~42 μm. Using $\lambda_0 = 800$ nm and $\beta_1 = 1/c$, Eq. (S4) yields $\beta_{\text{STD}} = 0.613$ fs/cm. Put this value into Eq. (1) in the main text, the phase-matching wavelength can be estimated to be ~290 nm, in striking agreement with the experimental result, see Fig. 1b of the main text.

# Supplementary note 3: Numerical simulation model

In this note, we describe the numerical model of this work. The electric field inside the HCF is assumed to preserve cylindrical symmetry and can be expanded as a linear superposition of $HE_{1m}$ modes: $\tilde{E}(z,r,\omega) = \sum_m \tilde{E}^m(z,\omega) J_0[u_{1m}(r/a)]$, where $\tilde{E}(z,r,\omega)$ is the electric field in the frequency domain, $z$ is the propagation distance, $\omega$ is the angular frequency of the pulse, $\tilde{E}^m(z,\omega)$ denotes the amplitude of the $HE_{1m}$ mode, $J_0$ is the zeroth-order Bessel function, $u_{1m}$ is the m[th] zero of $J_0$, $r$ is the radial coordinate, and $a$ is the core radius of the HCF. The numerical simulations presented in Figs. 2-6 of the main text are based on the multi-mode unidirectional pulse propagation equation[2,3], which incorporates a photoionization term and reads[2,3]:

$$\frac{\partial \tilde{E}^m(z,\omega)}{\partial z} = i\left(\beta^m(\omega) - \frac{\omega}{v_g}\right)\tilde{E}^m(z,\omega) - \frac{\alpha^m(\omega)}{2}\tilde{E}^m(z,\omega) + i\frac{\omega^2 \tilde{P}_{NL}^m(z,\omega)}{2c^2\varepsilon_0\beta^m(\omega)} \quad (S5)$$

where $\beta^m$ is the propagation constant of the $HE_{1m}$ mode, $\alpha^m(\omega)$ is the linear capillary loss of that mode, $\varepsilon_0$ is the vacuum permittivity, $\tilde{P}_{NL}^m(z,\omega)$ represents the nonlinear response in the frequency domain at the $HE_{1m}$ mode, which can be expressed as $\tilde{P}_{NL}^m(z,\omega) = 2\pi \int_0^a r dr J_0(u_{1m}r/a)\tilde{P}_{NL}(z,r,\omega)$. Here, $\tilde{P}_{NL}(z,r,\omega)$ is the nonlinear polarization in the frequency domain, which can be given as[4-6]:

$$\tilde{P}_{NL}(z,r,\omega) = F\left[\varepsilon_0 \chi^{(3)} E(z,r,t)^3 + P_{ion}(z,r,t)\right] \quad (S6)$$

where $F$ represents the Fourier transform, $\chi^{(3)}$ is the third-order nonlinear susceptibility associated with Kerr effect, $E$ is the electric field in the time domain, and $t$ is the time in a reference frame moving at the group velocity. $P_{ion}(z,r,t)$ is the ionization-induced plasma response given by[4-6]:

$$\frac{\partial P_{ion}(z,r,t)}{\partial t} = \frac{I_P}{E(z,r,t)}\frac{\partial \rho(z,r,t)}{\partial t} + \frac{e^2}{m_e}\int_{-\infty}^t \rho(z,r,t')E(z,r,t')dt' \quad (S7)$$

where $I_P$ is the ionization potential of the gas, $\rho$ is the plasma density, $e$ is the electronic charge, and $m_e$ is the electron mass. The plasma density evolves according to:

$$\frac{\partial \rho}{\partial t} = W(I)(\rho_{nt} - \rho) + \frac{s}{I_P}\rho I \quad (S8)$$

where $W(I)$ is the field-dependent ionization rate, $I$ is the laser pulse intensity, $\rho_{nt}$ is the neutral gas density, and $s$ is the cross section for collisional ionization.

The propagation constant $\beta^m$ in Eq. (S4) follows from the capillary dispersion relation[7]:

$$\beta^m(\omega) = \frac{\omega}{c}\sqrt{n_{gas}^2(\omega) - \frac{u_{1m}^2 c^2}{\omega^2 a^2}} \qquad (S9)$$

where $n_{gas}$ is the refractive index of the filling gas, which depends on light frequency, gas pressure, and temperature. The linear loss of the HCF can be given as[8]:

$$\alpha^m(\omega) = \frac{u_{1m}^2 c^2}{\omega^2 a^3} \frac{\eta^2 + 1}{\sqrt{\eta^2 - 1}} \qquad (S10)$$

where $\eta$ is the ratio of the refractive index of the cladding medium to that of the filling gas.

In the simulation, we used ~40-fs (full width at half maximum), 800-nm, Gaussian-shape pulses as input, and considered that it is coupled entirely into the HCF fundamental mode.

# Supplementary note 4: Experimental set-up

The high-energy multi-mode ultraviolet (UV) pulse generation setup is illustrated in Fig. S1. The system was driven by a commercial Ti: sapphire femtosecond laser (Legend Elite, Coherent), which delivered pulses at a central wavelength of 800 nm, with a pulse energy of 5 mJ, a duration of approximately 40 fs, and a repetition rate of 1 kHz. The driving pulses were first passed through a half-wave plate (HWP) and a thin-film polarizer (TFP), which together functioned as tunable attenuator for energy adjustment. The attenuated pulses were then coupled into a hollow capillary fibre (HCF) using concave mirrors with focal lengths of 2 m or 3 m. Two HCF configurations were employed: one with a core diameter of 500 µm and a length of 0.4 m, and the other with an 800 µm core diameter and a 1 m length. The input and output ends of the HCF were sealed with a 0.5-mm-thick coated fused silica (FS) window and a 1-mm-thick uncoated magnesium fluoride ($MgF_2$) window, respectively. The 0.4-m-long HCF was filled with either 9.3 bar of Ne or 3 bar of Ar gas, while the 1-m-long HCF was filled with 6.5 bar of He gas. The measured transmission efficiency for both configurations was approximately 80%. Finally, the generated multimode UV pulses at the HCF output were fully characterized through spectral analysis, pulse energy measurement, and beam profile diagnostics.

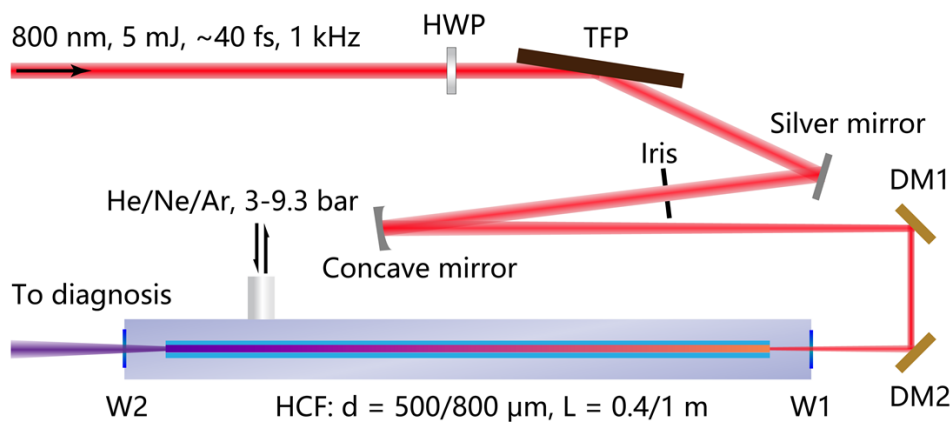

**Supplementary Fig. S1. Experimental set-up.** HWP, half-wave plate; TFP, thin film polarizer; DM1-DM2, dielectric mirrors; W1-W2, the first (W1) is the fused silica (FS) window, and the last (W2) is the magnesium fluoride ($MgF_2$) window; HCF, hollow capillary fibre; d, fibre diameter; L, fibre length.

# Supplementary note 5: Energy scaling of MMCR

As delineated in the main text, the energy level of the MMCR process can be scaled through appropriate adjustment of relevant system parameters. Figure S2 illustrates the spectral evolution dynamics as a function of increasing pump energy for two distinct experimental configurations. Despite exhibiting an order-of-magnitude difference in their respective energy scales, both configurations demonstrate qualitatively similar spectral behavior, consistently generating multimode UV pulses centered at approximately 300 nm, with measured pulse energies of 7.7 µJ and 113 µJ, respectively. For a more detailed characterization of the corresponding spectral profiles, please refer to Figs. 6(c) and 6(d) in the main text.

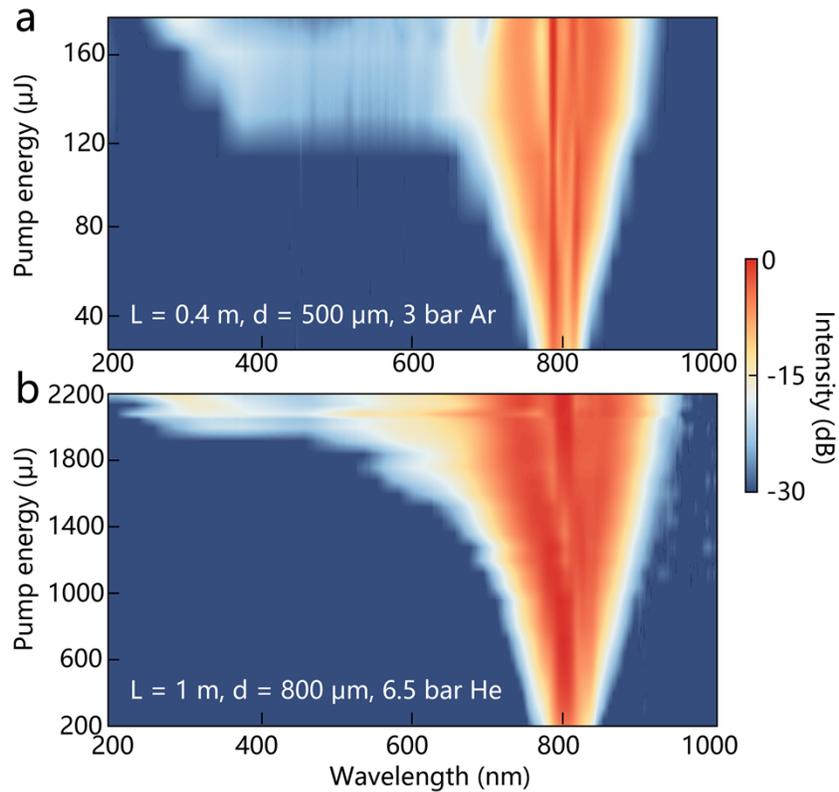

**Supplementary Fig. S2. Measured spectral evolution with different capillary geometries.** Evolution of the pulse spectra as a function of pump pulse energy for: (a) 0.4-m-long, 500-µm-diameter capillary fibre filled with 3 bar Ar gas, (b) 1-m-long, 800-µm-diameter capillary fibre filled with 6.5 bar He gas. The pump pulse parameters are the same as those used in Figs. 6(c) and 6(d) of the main text.

# Supplementary note 6: Conventional soliton-driven high-energy UV Cherenkov radiation

Although Cherenkov radiation via one-dimensional soliton dynamics in gas-filled HCFs is widely regarded as an ideal scheme for generating microjoule-level, tunable ultrafast UV pulses, scaling such sources to the tens or hundreds of microjoules remains a challenge. A principal difficulty lies in energy scaling while avoiding disruptive spatial nonlinear effects. To this end, the fibre core diameter must be enlarged quadratically, and the gas pressure reduced proportionally, to maintain the anomalous dispersion required for stable soliton propagation. This scaling, however, directly leads to a proportional extension of both the soliton fission length and the necessary HCF length[2].

For instance, simulations of conventional Cherenkov radiation in a 500-μm-core-diameter HCF filled with 0.77 bar He gas, pumped by a 40 fs, 550 μJ, 800 nm pulse, indicate that generating a ~300 nm, 45 μJ UV pulse would require a 40-m fibre length—impractical for most laboratories (see Fig. S3a). Pre-compressing the pump pulse to ~10 fs can shorten the required fibre length[2], yet such approaches impose significant constraints on the experimental footprint while substantially increasing system complexity and cost. As illustrated in Fig. S3b, even with a 10 fs, 308 μJ pump pulse, an 8-m fibre is still needed—far longer than the 0.4-m or 1-m HCFs employed in this work.

Scaling soliton dynamics to even higher energies, targeting hundred-microjoule UV Cherenkov radiation, would demand fibre lengths of approximately 100 m (with a 40 fs pump) or 20 m (with a 10 fs pump), as shown in Figs. S3c and S3d. In contrast, the MMCR scheme demonstrated here offers a significantly more compact and practical route to realizing high-energy ultrafast UV sources.

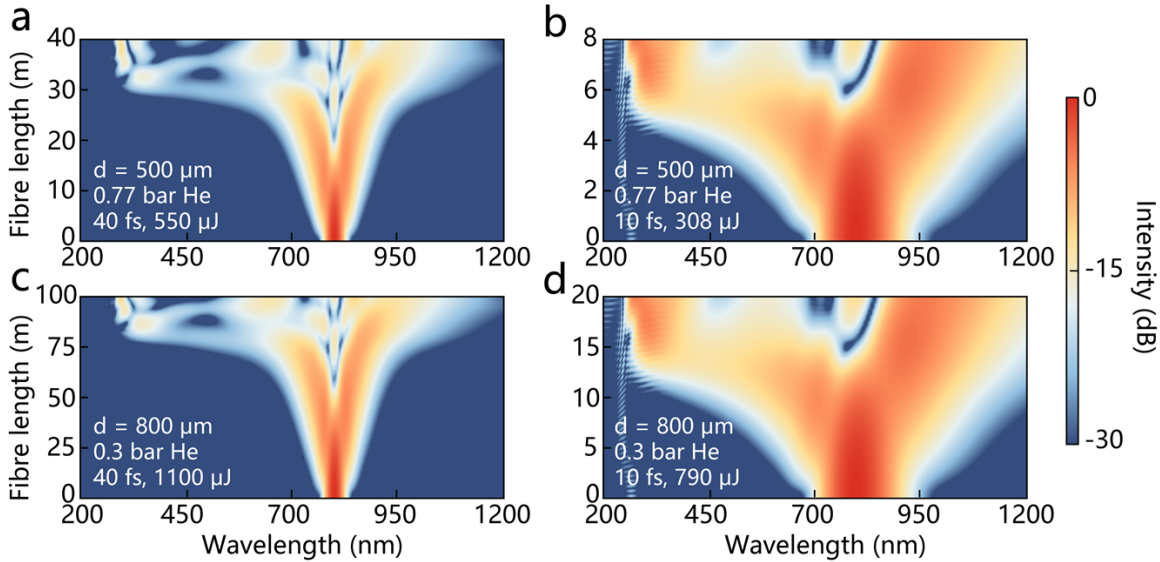

**Supplementary Fig. S3. Simulation of conventional soliton-driven Cherenkov radiation.** Simulated spectral evolution as a function of fibre length with different pump pulses and capillary geometries: (a) 500-μm-diameter capillary fibre pumped with 40 fs, 550 μJ at 0.77 bar He gas, (b) 500-μm-diameter capillary fibre pumped with 10 fs, 308 μJ at 0.77 bar He gas, (c) 800-μm-diameter capillary fibre pumped with 40 fs, 1100 μJ at 0.3 bar He gas, (d) 800-μm-diameter capillary fibre pumped with 10 fs, 790 μJ at 0.3 bar He gas. The simulation parameters were guided by the experimental results from Ref. [2] and the scaling laws of soliton dynamics, yielding a ~300 nm, hundred-microjoule-level deep-UV pulse.

## Supplementary References